\documentclass[12pt,final,a4paper]{article}
\usepackage{amsmath}
\usepackage{amssymb}

\input{tcilatex}
\input tcilatex

\begin{document}

\author{M. Abu-Shady \\
{\small Department of applied Mathematics, Faculty of Science, Menoufia
University, Egypt}}
\title{Mid-Point Technique for Calculating Divergent integrals}
\maketitle

\begin{abstract}
A mid-point technique is suggested to overcome the difficulties in other
techniques. The modified effective interaction quark potential which uses to
calculate different properties of the NJL model such as the constituent
quark mass, the pressure, and the energy density is solved using the present
technique. The present method gives good accuracy for the mathematical
problem and avoids the physical difficulty in the previous works.
\end{abstract}

\section{Introduction}

The appearance of divergent integrals in the different mathematical and
physical models is a real problem. This problem represents that the
treatment of divergence integrals did not give a unique value for divergent
integrals such as $\left[ 1\right] $. There are different methods for
dealing with the divergent integrals such as the analytic continuation [2],
by regularization [3], by summability methods [4], and by finite part
integrals [5, 6], and others [7, 8]. In these methods, the missing terms
have appeared so that no interpretation gives the exact results.

In this work, we focus on the quark models that deal with the strong
interactions between quarks inside nucleon. There are different quark models
such as the quark sigma model and its extension $\left[ 9-15\right] $ and
Nambu and Jona-Lasinio model (NJL) its extension $\left[ 16-19\right] $. The
NJL is taken as a model for the divergent integral $\left[ 16\right] $. The
NJL model has the divergent integral that obtained from Dirac sea to
overcome on this problem. Pauli-Villars's method and the three-momentum
cutoff methods are applied such as $\left[ 20,21\right] $.

Therefore, the N-midpoint technique is applied for the treatment of
divergence integrals in the NJL. In addition, the thermodynamic properties
are calculated in the framework of NJL model, in which the coupling constant
is generalized as a function of temperature and chemical potential that is
not considered in Ref. $\left[ 19\right] $

The paper is arranged as follows: In Sec. 2, the NJL model is briefly
written. Sec. 3, thermodynamic properties. are calculated. Sec. 4, the
discussion of results is explained.

\section{The \textbf{Nambu-Jona-Lasinio Model}}

The interactions between quarks $(q)$ via\ scalar $\bar{q}q$ and
pseudoscalar $\bar{q}\gamma _{5}\mathbf{\tau }q$ [18] in the framework of
NJL model is given through the Lagrangian density is, 
\begin{equation}
L\left( x\right) =\bar{q}(i\partial _{\mu }\gamma ^{\mu }-m_{0})q+\frac{G_{0}%
}{2}\left( (\bar{q}q)^{2}+(\bar{q}\gamma _{5}\mathbf{\tau }q)^{2}\right) , 
\tag{1}
\end{equation}%
$m_{0}$ is the bare quark mass. The $\mathbf{\tau }$ are the Pauli matrices.
The NJL Lagrangian satisfied the chiral symmetry property at $m_{0}$ $=0$.
In the present work, the coupling constant $G$ is extended to finite
temperature $\left( T\right) $and chemical potential $\left( \mu \right) $
as follows

\begin{equation}
G\left( T,\mu \right) =G_{0}e^{-A\left( T+\mu \right) }  \tag{2}
\end{equation}%
where, A is a free parameter\ in unit of MeV$^{-1}$or GeV$^{-1}$ units. The $%
G\left( T,\mu \right) $ tends to $G_{0}$ at $T=0$ and $\mu =0$

\section{The Used Method}

The N-midpoint technique is used to calculate the thermodynamic properties.
\ The scalar density is written as in Ref. $\left[ 18\right] $%
\begin{equation}
\rho _{s}=\left\langle \bar{q}q\right\rangle =\frac{6M}{\pi ^{2}}%
\int_{0}^{\infty }\frac{1}{\sqrt{p^{2}+M^{2}}}\left( n_{q}\left( T,\mu
\right) +n_{\bar{q}}\left( T,\mu \right) -1\right) p^{2}dp.  \tag{3}
\end{equation}%
The dimensionless form of Eq. (3) is obtained as follows%
\begin{equation}
\rho _{s}=\left\langle \bar{q}q\right\rangle =\frac{6M}{\pi ^{2}}^{\prime
}\int_{0}^{\infty }\frac{\left( n_{q}\left( T^{\prime },\mu ^{\prime
}\right) +n_{\bar{q}}\left( T^{\prime },\mu ^{\prime }\right) -1\right) }{%
\sqrt{p^{\prime 2}+M^{\prime 2}}}p^{^{\prime }2}dp^{\prime },  \tag{4}
\end{equation}%
where%
\begin{equation}
p^{\prime }=\frac{p}{f_{\pi }},T^{\prime }=\frac{T}{f_{\pi }},M^{\prime }=%
\frac{M}{f_{\pi }}\text{, and }\mu ^{\prime }=\frac{\mu }{f_{\pi }}.  \tag{5}
\end{equation}%
$f_{\pi }=93$ MeV is pion decay constant. The third term in Eq. (3) is a
divergent term. To apply the mid-point method, the formula $y=e^{-p^{\prime
}}$ is used at $\left\{ y=1\text{ at }p^{\prime }=0\text{ and }y=0\text{ at }%
p^{\prime }=\infty \right\} $. Therefore, $\rho _{s}$ is calculated as a
function of $y$ as follows:

\begin{equation}
\rho _{s}=\frac{6M}{\pi ^{2}}^{\prime }\int_{0}^{1}\frac{\left( n_{q}\left(
T^{\prime },\mu ^{\prime }\right) +n_{\bar{q}}\left( T^{\prime },\mu
^{\prime }\right) -1\right) }{y\sqrt{\ln ^{2}y+M^{^{\prime }2}}}\ln ^{2}ydy,
\tag{6}
\end{equation}%
where%
\begin{equation}
n_{q}\left( T^{\prime },\mu ^{\prime }\right) =\frac{1}{1+\exp [\frac{(\sqrt{%
\ln ^{2}y+M^{^{\prime }2}}-\mu ^{\prime })}{T^{\prime }}]},\ \ \ \ \ \ \ n_{%
\bar{q}}\left( T^{\prime },\mu ^{\prime }\right) =n_{q}\left( T^{\prime
},-\mu ^{\prime }\right) .  \tag{7}
\end{equation}
The analytic function of $\rho _{s}$ is calculated by using the N-midpoint
method as in Ref. [23] as follows:

\begin{equation*}
\rho _{s}=\frac{6M^{\prime }}{n\ \pi ^{2}}\sum_{i=0}^{n-1}\frac{\ln
^{2}\left( A_{i}\right) }{A_{i}\sqrt{M^{^{\prime }2}+\ln ^{2}\left(
A_{i}\right) }}[\frac{1}{\exp \left( \frac{1}{T^{\prime }}\left( \mu
^{\prime }+\sqrt{M^{^{\prime }2}+\ln ^{2}\left( A_{i}\right) }\right)
\right) +1}+
\end{equation*}

\begin{equation}
+\frac{1}{\exp \left( \frac{1}{T^{\prime }}\left( \sqrt{M^{^{\prime }2}+\ln
^{2}\left( A_{i}\right) }-\mu ^{\prime }\right) \right) +1}-1],  \tag{8}
\end{equation}%
where $A_{i}$ is defined as follows: 
\begin{equation}
A_{i}=\frac{1}{n}i+\frac{1}{2n}.  \tag{9}
\end{equation}%
Similarly, the thermodynamic potential is written as a function of variable $%
y$ using $y=e^{-p^{\prime }}$ 
\begin{eqnarray}
\Omega &=&\frac{\left( M^{\prime }-m_{0}^{\prime }\right) ^{2}}{2G^{\prime }}%
-\frac{6}{\pi ^{2}}\int_{0}^{1}\frac{\ln ^{2}y}{y}dy\{\sqrt{\ln
^{2}y+M^{^{\prime }2}}+  \notag \\
&&T~^{\prime }\ln [1+\exp \left( -\frac{\sqrt{\ln ^{2}y+M^{^{\prime }2}}+\mu
^{\prime })}{T^{\prime }}\right) ]+  \notag \\
&&+T~^{\prime }\ln [1+\exp \left( -\frac{(\sqrt{\ln ^{2}y+M^{^{\prime }2}}%
-\mu ^{\prime })}{T^{\prime }}\right) ]\},  \TCItag{10}
\end{eqnarray}%
where%
\begin{equation}
m_{0}^{\prime }=\frac{m_{0}}{f_{\pi }}\text{ and }G^{\prime }=G~f_{\pi }^{2}.
\tag{11}
\end{equation}
The analytic form of $\Omega $ is 
\begin{eqnarray}
\Omega (T^{\prime },\mu ^{\prime }) &=&\frac{\left( M^{\prime
}-m_{0}^{\prime }\right) ^{2}}{2G^{\prime }}-\frac{6}{n\ \pi ^{2}}%
\sum_{i=0}^{n-1}\frac{1}{A_{i}}\left( \ln ^{2}\left( A_{i}\right) \right) (%
\sqrt{M^{\prime 2}+\ln ^{2}\left( A_{i}\right) }+  \notag \\
&&T^{\prime }\ln \left( \exp \left( -\frac{1}{T^{\prime }}\left( \mu
^{\prime }+\sqrt{M^{^{\prime }2}+\ln ^{2}\left( A_{i}\right) }\right)
\right) +1\right) +  \notag \\
&&+T^{\prime }\ln \left( \exp \left( -\frac{1}{T^{\prime }}\left( -\mu
^{\prime }+\sqrt{M^{\prime 2}+\ln ^{2}\left( A_{i}\right) }\right) \right)
+1\right) ).  \TCItag{12}
\end{eqnarray}%
Also, we can define the energy density as in Ref. [25] as follows:

\begin{equation}
E(T^{\prime },\mu ^{\prime })=-P(T^{\prime },\mu ^{\prime })+T^{\prime }%
\frac{\partial P(T^{\prime },\mu ^{\prime })}{\partial T^{\prime }}+\mu
^{\prime }\frac{\partial P(T^{\prime },\mu ^{\prime })}{\partial \mu
^{\prime }},  \tag{13}
\end{equation}%
where 
\begin{equation}
P(T^{\prime },\mu ^{\prime })=-\Omega (T^{\prime },\mu ^{\prime }).  \tag{14}
\end{equation}

\bigskip

\section{Results and Discussion}

In this section, we give the features of the midpoint technique in
comparison with other methods. The fewer errors are obtained in comparison
with other numerical methods as in Ref. $\left[ 23\right] $. In comparison
with the cut-off technique. Two advantages are found, from the mathematical
view, the upper of momentum value depends on to the parameter $n$, which
gives a good accuracy by increasing parameter $n$ as noted in Table (1). \
Another advantage, by transforming the infinite divergence integral to
finite divergence, we control in the missing terms that found in other
methods. From the physical situation, the color superconductivity is found
at the critical chemical potential that is the order of the cut-off
parameter. Thus, the regularization procedure has a small effect on the
analysis of the color superconductivity $\left[ 11\right] $.

\textbf{Table (1)}. \ The relation between accuracy number $n$, the maximum
value of momentum $p$, and Sea term in Eq. (8) $\left( \text{Third Term}%
\right) $

\begin{tabular}{|l|l|l|l|l|}
\hline
${\small n}$ & {\small 400} & {\small 600} & {\small 800} & {\small 1000} \\ 
\hline
{\small Momentum} & $621.\,\allowbreak 67${\small \ MeV} & $%
659.\,\allowbreak 38${\small \ MeV} & $\allowbreak 686.\,\allowbreak 13$%
{\small \ MeV} & $706.\,\allowbreak 88${\small \ MeV} \\ \hline
{\small Sea Term }$\times 10^{13}$ & $2.\,\allowbreak 740\,7$ {\small MeV} & 
$\allowbreak 4.\,\allowbreak 862\,3$ {\small MeV} & $7.\,\allowbreak 265\,6$ 
{\small MeV} & $9.\,\allowbreak 894\,6$ {\small MeV} \\ \hline
\end{tabular}

\ \ \ \ \ \ \ \ 

\ In Fig. 1, the coupling constant is plotted as a function of temperature
by using Eq. (2). At zero temperature and baryon chemical potential, the
running coupling constant tends to $G_{0}$ in the vacuum, then by increasing
temperature, one notes that the coupling constant decreases with increasing
temperature. In addition, the coupling constant is strongly affected with
the baryon chemical potential at higher values of temperatures. By
increasing baryon chemical, the running coupling drops to lowers values. In
Fig. 2, the constituent quark mass is plotted as a function temperature. One
notes that the constituent quark mass is steady function up to 75 MeV, then
the constituent quark mass decreases with increasing temperature at zero
chemical potential. By increasing chemical potential, the constituent quark
mass drops to lower values. In comparison with Refs. $\left[ 18,19\right] $,
the behavior of constituent quark mass is the qualitative agreement with the
results of Refs. $\left[ 18,19\right] $, in which the running coupling is
not considered. In Fig. 3, the pressure is plotted as a function of
temperature. One notes that the pressure is increasing function with
temperature and shifts to higher values by increasing baryon chemical
potential. Moreover, the pressure is not sensitive up to 100 MeV, then the
pressure is a sensitive quantity when baryon chemical potential increases
more than 150 MeV. This finding is agreement with the conclusion of Refs. $%
\left[ 24,25\right] $. A similar situation for energy density, in which the
energy density increases with increasing temperature and shifts to higher
values by increasing baryon chemical potential. In addition, the energy
density is steady function up to 75 MeV.

\ In Fig. 5, the consistent quark mass is plotted in the two cases. In the
first case, when the coupling constant is constant and the second case is
depended on temperature and baryon chemical potential. The consistent quark
mass drops to lower values at higher values of temperature in the second
case. A similar situation, the pressure drops to lower values when the
dependent coupling temperature is considered. Therefore, the effect of
running coupling constant is not qualitatively changed which leads to the
phase transition is kept as crossover as in Refs. $\left[ 18,19\right] $
\bigskip 

\ \ \ \ \ 

\section{Conclusion}

In this work, \ the mid-point method is employed to carry out the divergence
integral in the NJL model that found from Dirac sea. To apply this method,
the infinite divergence integral to finite divergence integral is
transformed. Therefore, we avoid the missing terms that found in the
previous works such as the cut-off technique. In cut-off technique that is
applied in many works has disadvantages, in which all observables depend on
the choice of cut-off parameter. In the present method, the calculated
integral depends on the accuracy parameter $\left( n\right) $ only. Thus,
good accuracy is obtained by controlling in the accuracy parameter $\left(
n\right) .$ The observables such as constituent quark mass, pressure, and
energy density are calculated by using the present method with a good
accuracy.

The coupling constant is extended to finite temperature and baryon chemical
potential. The suggested running coupling constant decreases with increasing
temperature and baryon chemical potential. In addition, the running coupling
constant tends to the normal coupling constant at zero temperature and
baryon chemical potential. All observables are dropped to lower values when
the running coupling constant is included. Therefore, the coupling constant
is not affected on qualitative behavior of the present observables. This
finding is not considered in the previous works such as $\left[ 16-19\right]
.$

\section{Acknowledgments}

The author thanks the reviewers and editor for their constructive
suggestions to improve the quality of this paper.

\end{document}